# The effect of atmospheric turbulence on entangled orbital angular momentum states

**C Gopaul[1] and R Andrews[1]**
[1]Department of Physics, The University of the West Indies, St. Augustine, Trinidad and Tobago

E-mail: cgopaul@mysta.uwi.edu ; randrews@fsa.uwi.tt;

**Abstract.** We analyze the effect of atmospheric Kolmogorov turbulence on entangled orbital angular momentum states generated by parametric down-conversion. We calculate joint and signal photon detection probabilities and obtain numerically their dependence on the mode-width-to-Fried-parameter ratio. We demonstrate that entangled photons are less robust to the effects of Kolmogorov turbulence compared to single photons. In contrast, signal photons are more robust than single photons in the lowest-order mode. We also obtain numerically a scaling relation between the value of the mode-width-to-Fried-parameter ratio for which the joint detection probabilities is a maximum and the momentum mismatch between signal and idler photons after propagation through the medium.

## 1. Introduction

Quantum communication utilizing the phenomenon of entanglement has many applications such as quantum teleportation [1], quantum cryptography [2, 3] and superdense coding [4]. Multi-dimensional entangled orbital angular momentum (OAM) states can be utilized to generate arbitrary base-N quantum digits thereby allowing the implementation of higher capacity optical communication systems. This feature has generated considerable interest with respect to encoding quantum and classical information [2, 5]. In addition it has been shown that the (OAM) of photons can be used to encode data onto a laser beam for transmitting information in free-space optical systems [6]. The performance of free-space optical communication systems is however severely degraded by decoherence caused by atmospheric turbulence in the optical channel, resulting in fluctuations in both the intensity and phase of the received signal. Recently, C. Paterson [7] estimated the theoretical performance limits for communication links based on the OAM of single photons with no entanglement. It was demonstrated that for Kolmogorov atmospheric statistics this system is significantly affected, even in the weak turbulence regime, with narrower and lower-order modes being more robust. Spontaneous parametric down-conversion generates pairs of down-converted photons entangled in OAM [5]. It has been demonstrated that Laguerre-Gaussian (LG) modes possess well defined orbital angular momemta [8]. In the paraxial approximation, these modes are the eigenstates of the orbital angular momentum operator with eigenvalue equal to $l\hbar$ per photon. Xi-Fen Ren et al [9] calculated the superposition coefficients of down converted correlated LG modes and showed that this probability amplitude decreases almost exponentially with increasing OAM. In this paper we investigate the effects of Kolmogorov turbulence on entangled OAM photon states by calculating the joint detection probability of entangled LG modes. We compare the effects of Kolmogorov turbulence on entangled photon pairs and single photons.

## 2. Theory

The two-photon quantum state in the Laguerre-Gaussian (LG) basis is given by [10]

$$|\Psi\rangle = \sum_{l_1,p_1 l_2,p_2} C^{l_1,l_2}_{p_1,p_2} |l_1,p_1; l_2,p_2\rangle \qquad (1)$$

where $(l_1, p_1)$ represent the signal mode numbers and $(l_2, p_2)$ represent the corresponding idler mode numbers, with $l_1 + l_2 = l_0$, where $l_0$ is the OAM of the pump. For a cw pump beam and collinear phase matching of signal and idler photons, the probability amplitude $C^{l_1,l_2}_{p_1,p_2}$ is given by [11, 12]

$$C^{l_1,l_2}_{p_1,p_2} \sim \int dr_\perp \Phi(r_\perp) \left[LG^{l_1}_{p_1}(r_\perp)\right]^* \left[LG^{l_2}_{p_2}(r_\perp)\right]^* \qquad (2)$$

where $\Phi(r_\perp)$ is the transverse spatial profile of the pump beam at the input face of the crystal (z=0) and $LG^l_p(r_\perp)$ is an LG mode in the z=0 plane. The normalized Laguerre-Gaussian mode at its beam waist (z=0) is given by

$$LG^l_p(r,\theta) = R^l_p(r) \frac{\exp(il\theta)}{\sqrt{2\pi}} \qquad (3)$$

where

$$R^l_p(r) = 2\left(\frac{p!}{(|l|+p)!}\right)^{1/2} \frac{1}{w_0} \left(\frac{r\sqrt{2}}{w_0}\right)^{|l|} L^{|l|}_p\left(\frac{2r^2}{w_0^2}\right) \exp\left(-\frac{r^2}{w_0^2}\right) \qquad (4)$$

with normalization constant $\left(\frac{2p!}{\pi(|l|+p)!}\right)^{1/2}$, and

$$L^{|l|}_p(x) = \sum_{m=0}^{p} (-1)^m \frac{(|l|+p)!}{(p-m)!(|l|+m)!m!} x^m \qquad (5)$$

$w_0$ is the mode width, p is the number of nonaxial radial nodes of the mode, $l$ is the OAM number. Atmospheric turbulence scatters the beam as it propagates, producing phase aberrations that perturb the entangled modes. In the weak turbulence regime, this can be considered as a pure phase perturbation $\phi(r_\perp)$. In the presence of turbulence the probability amplitude (2) becomes

$$\tilde{C}^{l_1,l_2}_{p_1,p_2} \sim \int dr_\perp \Phi(r_\perp) \left[LG^{l_1}_{p_1}(r_\perp)\right]^* \left[LG^{l_2}_{p_2}(r_\perp)\right]^* \exp[i\phi(r_\perp)] \qquad (6)$$

## 3. Joint detection probabilities

The conditional probability of coincidence of photons with eigenvalues $l_1$ in the signal mode and $l_2$ in the idler mode is given as $P(l_1,l_2|\Psi) = \sum_{p_1,p_2} \left|\tilde{C}^{l_1,l_2}_{p_1,p_2}\right|^2$. Since the aberrations are random, taking the ensemble average gives the joint detection probability $P(l_1,l_2)$ that a measurement of the OAM of the signal photon and the idler photons yields values $l_1$ and $l_2$ respectively. This probability is defined as

$$P(l_1,l_2) = \left\langle \sum_{p_1,p_2} \left|\tilde{C}^{l_1,l_2}_{p_1,p_2}\right|^2 \right\rangle \qquad (7)$$

where $\langle \rangle$ represents an ensemble average. For illustrative purposes we consider the mode subspace where $p_1 = p_2 = 0$. Assuming the statistics of the Kolmogorov turbulence aberrations to be isotropic and a pump beam in a pure LG mode with mode numbers $(l_0, p_0)$ we obtain

$$P(l_1, l_2) \propto \iiint \left[R_{p_0}^{l_0}(r')\right]^* \left[R_{p_0}^{l_0}(r)\right] \left[R_0^{l_1}(r')\right] \left[R_0^{l_1}(r)\right]^* \left[R_0^{l_2}(r')\right] \left[R_0^{l_2}(r)\right]^*$$
$$\times \langle \exp\{-i[\phi(r, \Delta\theta) - \phi(r', 0)]\}\rangle \exp[-i\Delta l \Delta\theta] r' dr' r dr d\Delta\theta \tag{8}$$

where, $\Delta\theta = \theta - \theta'$ and $\Delta l = l_1 + l_2 - l_0$. Assuming the phase fluctuations to be a Gaussian random process, so that $\langle \exp(ix) \rangle = \exp\left(-\frac{1}{2}\langle |x|^2 \rangle\right)$, we can write

$$\langle \exp\{-i[\phi(r, \Delta\theta) - \phi(r', 0)]\}\rangle = \exp\left(-\frac{1}{2}D_\phi(|\underline{r} - \underline{r}'|)\right) \tag{9}$$

such that $D_\phi(|\underline{r} - \underline{r}'|) = 6.88(|\underline{r} - \underline{r}'|/r_0)^{5/3}$ for Kolmogorov turbulence where $r_0$ is the Fried parameter [13].

## 4. Signal photon detection probabilities

The probability, $P(l_1)$, for finding one photon with eigenvalue $l_1$ in the signal mode is obtained from the joint detection probability $P(l_1, l_2)$ by summing over all idler modes. Thus,

$$P(l_1) = \sum_{l_2, p_2} P(l_1, l_2) \propto \iiint \left[R_{p_0}^{l_0}(r)\right]^* \left[R_{p_0}^{l_0}(r)\right] \left[R_{p_1}^{l_1}(r)\right] \left[R_{p_1}^{l_1}(r)\right]^*$$
$$\times \langle \exp\{-i[\phi(r, \Delta\theta) - \phi(r, 0)]\}\rangle \exp[-i\Delta l \Delta\theta] r dr d\Delta\theta \tag{10}$$

where in (10) $\Delta l = l_1 - l_0$.

## 5. Results

We substitute (9) into (8) and evaluate the integrals in (8) numerically. In figure 1(a) we consider the probabilities for OAM measurements $P(l)$ with single photons [7]. $l_0$ is the initial OAM, $l$ is the final OAM, the initial and final mode indices $p_0 = p = 0$ and $w_0$ is the LG mode width. We plot $P(l)$ against the ratio of $w_0$ to the Fried parameter $(r_0)$ for the case where momentum is conserved $(\Delta l = 0)$. The results show that in the case of single photon scattering $P(l)$ decays more rapidly as $l_0$ increases. In figure 1(b) and figure 1(c) we consider the quantum entangled probability, $P(l_1, l_2)$, for entangled modes generated by a pump beam of width, $w_0$, with $p_0 = 0$ and signal and idler indices $p_1 = p_2 = 0$. Here we plot $P(l_1, l_2)$ against $w_0/r_0$ for the case where momentum is conserved $(\Delta l = 0)$ with $l_1 = 0$ and for increasing values of the pump and idler OAM. It is observed that the probabilities for the angular momentum measurements of entangled photons exhibit a similar behavior to single photons but there is a faster fall-off for all values of $l_0$. In figure 1(c) we examine a pump in the lowest-order LG mode ($l_0 = p_0 = 0$), and signal and idler OAM values $l_1 = -l_2 = |l|$ and $p_1 = p_2 = 0$. This corresponds to the situation in which both the pump as well as the signal/idler pairs have zero total angular momentum. We observe that $P(l_1, l_2)$ decays to zero more rapidly as the signal photon OAM increases. This shows that lower OAM signal and idler modes are more robust to turbulence. Figure 1(d) is a plot of signal photon probabilities, $P(l_1)$, versus ($w_0/r_0$) with $p_0 = p_1 = 0$ and pump OAM equal to signal OAM, i.e., $l_0 = l_1$. Qualitatively, the results in figure 1(d) are similar to those obtained in figure 1(a). However, $P(l_1)$ decays less rapidly than $P(l)$ for $l_0 = 0$. For the high-order modes $P(l_1)$ gives essentially the same decay profile as $P(l)$. This occurs since $P(l_1)$ depends on the spatial mode profile of the pump beam, whereas, for $P(l)$ the effect of any pump is absent. Our results show that signal photons are more robust to turbulence than unentangled single photons for $l_0 = 0$.

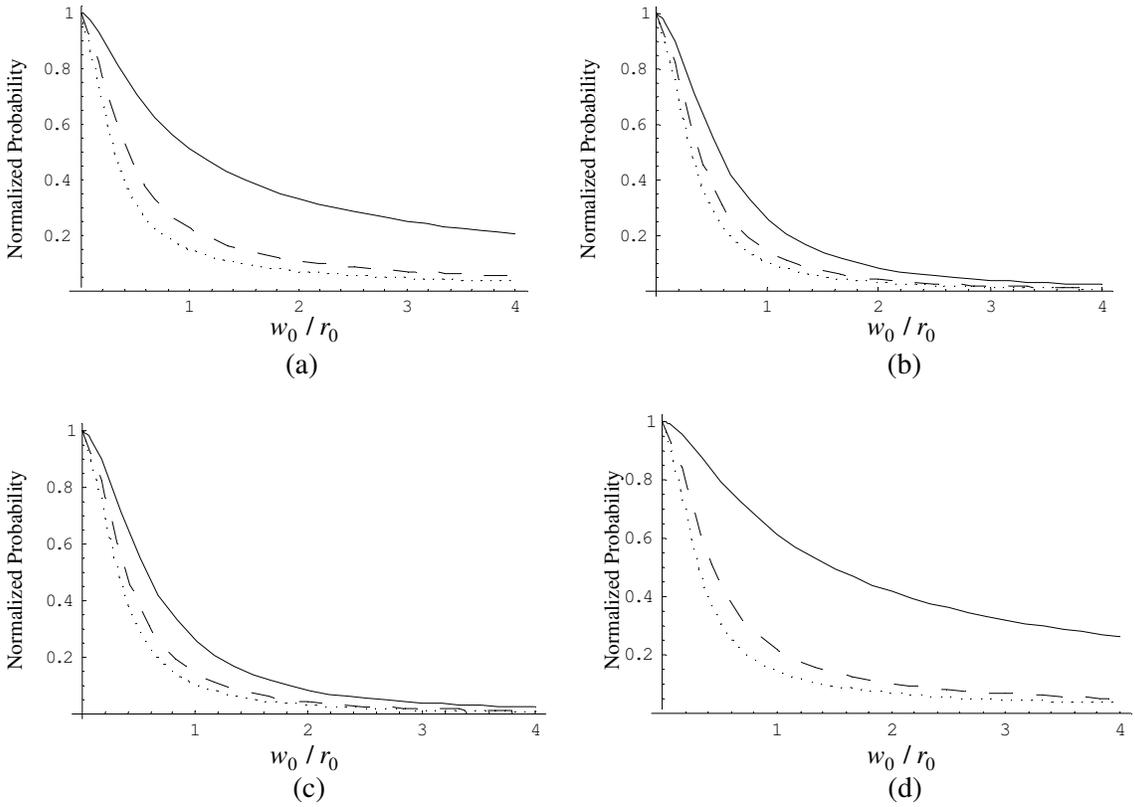

Figure 1: (a) Single photon detection probabilities $P(l)$ versus ($w_0/r_0$) with $p_0 = p = 0$ where $l_0 = l = 0$ (solid line); $l_0 = l = 1$ (dashed line) and $l_0 = l = 2$ (dotted line). (b) Joint detection probabilities $P(l_1, l_2)$ versus ($w_0/r_0$) for entangled modes with $p_0 = 0$ and $p_1 = p_2 = 0$ for given values of $l_0$, where $l_0 = l_2 = 0, l_1 = 0$ (solid line); $l_0 = l_2 = 1, l_1 = 0$ (dashed line) and $l_0 = l_2 = 2, l_1 = 0$ (dotted line). (c) Joint detection probabilities $P(l_1, l_2)$ versus ($w_0/r_0$) for entangled modes with $l_0 = p_0 = 0$ and $p_1 = p_2 = 0$ for particular values of $|l|$, where $|l| = 0$ (solid line); $|l| = 1$ (dashed line) and $|l| = 2$ (dotted line). (d) Signal photon probabilities $P(l_1)$ versus ($w_0/r_0$) with $p_0 = p_1 = 0$, where $l_0 = l_1 = 0$ (solid line); $l_0 = l_1 = 1$ (dashed line) and $l_0 = l_1 = 2$ (dotted line). In all cases $\Delta l = 0$.

In figure 2 we consider the case where angular momentum is not conserved $(\Delta l \neq 0)$ for single and entangled photons. In figure 2(a) we plot $P(l)$ against $w_0/r_0$ for single photons with $l_0 = p_0 = p = 0$. Here $P(l)$ increases from zero to a maximum and then decays. As $\Delta l$ increases the value of $w_0/r_0$ for which the probability is a maximum, $(w_0/r_0)_{max}$, also increases. In figure 2(b) we plot $P(l_1, l_2)$ against $w_0/r_0$ for entangled photons. Again we choose $l_0 = p_0 = 0$ and $l_2 = 0$, $p_1 = p_2 = 0$ such that $l_1 = \Delta l \neq 0$. For a given value of $\Delta l$ we observe a sharper increase from zero to a maximum and then a faster decay compared to that of single photons.

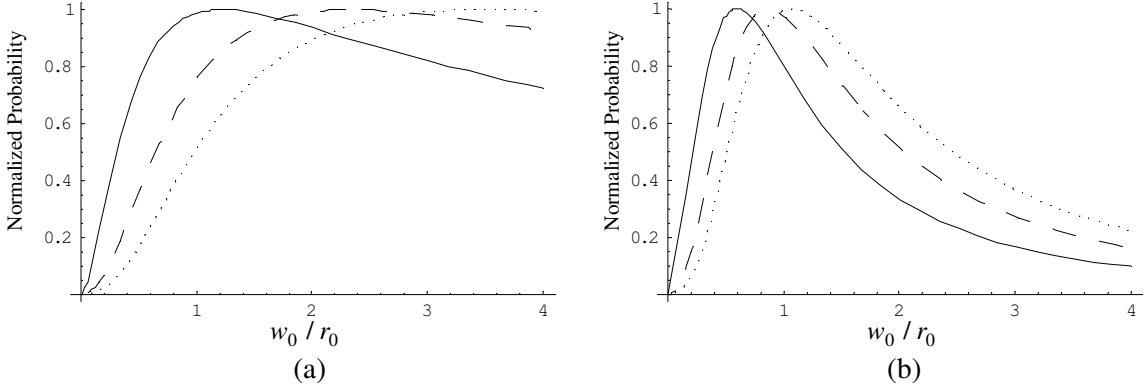

Figure 2: (a) Single photon detection probabilities $P(l)$ versus ($w_0/r_0$) with $l_0 = p_0 = 0$, $p = 0$ and $\Delta l \neq 0$, where $\Delta l = 1$ (solid line); $\Delta l = 2$ (dashed line) and $\Delta l = 3$ (dotted line). (b) Joint detection probabilities $P(l_1, l_2)$ versus ($w_0/r_0$) for entangled modes with $l_0 = p_0 = 0$, $p_1 = p_2 = 0$, $w_0 = w_p$ and $\Delta l \neq 0$, where $\Delta l = 1, l_1 = 1$ (solid line); $\Delta l = 2, l_1 = 2$ (dashed line) and $\Delta l = 3, l_1 = 3$ (dotted line).

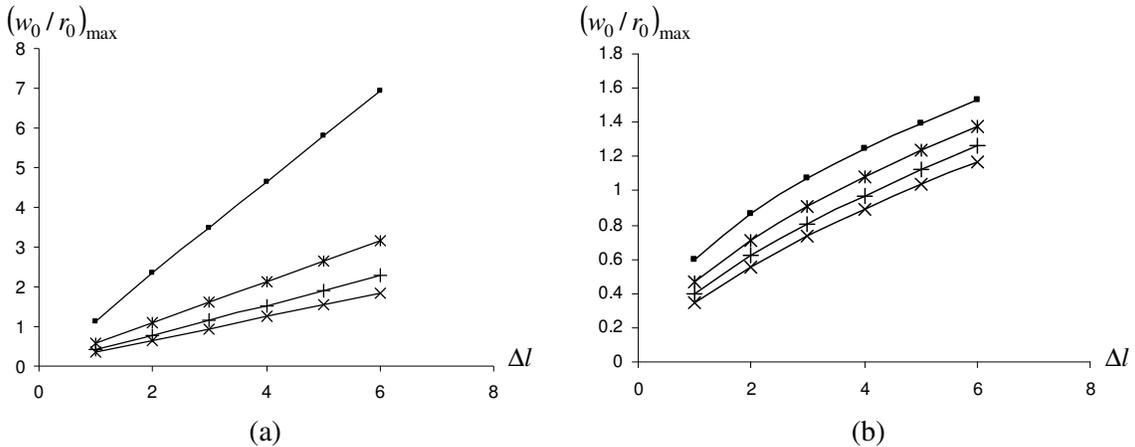

Figure 3: $(w_0/r_0)_{max}$ versus $\Delta l$ for (a) single photons where $l_0 = 0$ (···), $l_0 = 1$ (***), $l_0 = 2$ (+++) and $l_0 = 3$ (xxx). (b) entangled photons with $l_1 = \Delta l$, where $l_0 = l_2 = 0$ (···), $l_0 = l_2 = 1$ (***), $l_0 = l_2 = 2$ (+++) and $l_0 = l_2 = 3$ (xxx).

In figure 3 we plot $(w_0/r_0)_{max}$ against $\Delta l$ for single photons (a) and entangled photons with (b) for increasing values of $l_0$. We observe a linear relation between $(w_0/r_0)_{max}$ and $\Delta l$ for single photons similar to Wien's displacement law describing blackbody radiation. However for entangled photons the relation is nonlinear.

## 6. Conclusions

We have quantitatively described the effects of Kolmogorov atmospheric turbulence on the joint detection probabilities of entangled OAM states. We examined the two distinct cases where momentum is conserved and when there is a momentum mismatch after propagation through the turbulent medium. The results demonstrate that when photon pairs conserve momentum the joint detection probability decays rapidly as atmospheric turbulence increases. It is found that the joint detection probabilities of entangled photons with lower total OAM decay less rapidly than those with higher OAM values. Furthermore, entangled photons with smaller mode widths are more robust to turbulence. Similar qualitative results have been obtained by Smith and Raymer [14] using a two photon wave function approach. Our results demonstrate that entangled signal and idler photons are more susceptible to turbulence than single photons. This fact that can be used to aid in the design of quantum communication systems utilizing data encoded onto the OAM of entangled beams. In the case where momentum is not conserved the joint detection probability rises to maximum and decays more rapidly than for single photons. The value of the mode width to Fried parameter ratio for which the single photon probability is a maximum scales linearly with the momentum mismatch. In the case of entangled photons the scaling is not linear. One possible application of the measurement of the detection probabilities when momentum is not conserved is the determination of turbulence in free-space optical communication systems. A measurement of the joint detection probabilities of a fixed value of $\Delta l$ can be used to obtain $(w_0/r_0)_{max}$. Utilizing entangled photons for such measurements yields a more accurate value of the Fried parameter since the detection probabilities peak more sharply around $(w_0/r_0)_{max}$. Further investigation is required to determine the effect of anisotropic turbulence statistics on the joint detection probabilities.

We have also considered the effects of turbulence on the signal photon probabilities when momentum is conserved. It was found that the signal photon probabilities and unentangled single photon probabilities exhibited similar decay profiles except when $l_0 = 0$. In the special case of $l_0 = 0$, the signal photon probability decays more slowly with $(w_0/r_0)$ than the single photon probability. This shows that signal photons are more robust to the effects of turbulence than unentangled single photons. It indicates that information encoded onto the OAM of signal photons would be less susceptible to atmospheric scattering.